\documentclass[5p,sort&compress]{elsarticle}
\usepackage{hyperref}
\usepackage{etoolbox}
\makeatletter
    \patchcmd\@combinedblfloats{\box\@outputbox}{\unvbox\@outputbox}{}{\errmessage{\noexpand patch failed}}
    \patchcmd{\tnotemark}{\ding{73}}{\dag}{}{\@latex@error{Failed to path \string\tnotemark\space for \string\ding{73}}}
    \patchcmd{\tnotetext}{\ding{73}}{\dag}{}{\@latex@error{Failed to path \string\tnotetext\space for \string\ding{73}}}
\makeatother
\usepackage{graphicx}
\usepackage{subfigure}
\usepackage{xcolor}
\usepackage{placeins}

\usepackage{amsmath}
\usepackage{mathtools}
\usepackage{MnSymbol}
\usepackage{dsfont}
\usepackage{bbold}
\usepackage{slashed}

\usepackage{booktabs}
\usepackage{dcolumn}
\usepackage{widetable}

\usepackage[squaren]{SIunits}

\newcommand{\Ef}{E_{\vec{p}^\prime}}
\newcommand{\pbar}{\overline{p}}

\newcommand{\la}{\langle}
\newcommand{\ra}{\rangle}
\newcommand{\MS}{\overline{\mathrm{MS}}}
\newcommand{\PCACFF}{PCAC${}_{\mathrm{FF}}$}
\newcommand{\overbar}[1]{\mkern 1.5mu\overline{\mkern-1.5mu#1\mkern-1.5mu}\mkern 1.5mu}
\newcommand{\MSbar}{\overbar{\text{MS}}}
\begin{document}
\flushbottom
\begin{frontmatter}

\author[cor1,cor2]{G.~S.~Bali}
\author[cor1]{S.~Collins}
\author[cor1]{M.~Gruber}
\author[cor1]{A.~{Sch\"afer}}
\author[cor1]{P.~{Wein}}
\author[cor1]{T.~{Wurm}}
\ead{thomas.wurm@ur.de}
\address[cor1]{Institut f\"ur Theoretische Physik, Universit\"at Regensburg, 93040 Regensburg, Germany}
\address[cor2]{Department of Theoretical Physics, Tata Institute of Fundamental Research, Homi Bhabha Road, Mumbai 400005, India}

\title{Solving the PCAC puzzle for nucleon axial and pseudoscalar form factors\tnoteref{t1}}
\tnotetext[t1]{RQCD Collaboration}
\begin{abstract}%
    It has been observed in multiple lattice determinations of isovector axial and pseudoscalar nucleon form factors, that, despite the fact that the partial conservation of the axialvector current is fulfilled on the level of correlation functions, the corresponding relation for form factors (sometimes called the generalized Goldberger--Treiman relation in the literature) is broken rather badly. In this work we trace this difference back to excited state contributions and propose a new projection method that resolves this problem. We demonstrate the efficacy of this method by computing the axial and pseudoscalar form factors as well as related quantities on ensembles with two flavors of improved Wilson fermions using pion masses down to \unit{150}{\mega\electronvolt}. To this end, we perform the $z$-expansion with analytically enforced asymptotic behaviour and extrapolate to the physical point.
\end{abstract}
\end{frontmatter}

\section{Introduction}%
The axial nucleon structure is central for the description of weak interactions and plays a prominent role in long-baseline neutrino experiments, where it is important for a precise determination of the neutrino flux and the cross section for nuclear targets~\cite{AguilarArevalo:2010zc,McGivern:2016bwh,Katori:2016yel}. The nonperturbative information encoded in the corresponding form factors is manifold. For example, the isovector axial form factor is linked to the flavor asymmetry in the difference between helicity aligned and anti-aligned quark densities in impact parameter space~\cite{Burkardt:2002hr}. While the axial coupling $g_A$ is measured quite precisely in $\beta$-decay, the nucleon form factors that encode the spatial structure are much less well known. The axial form factor, $G_A(Q^2)$, and the induced pseudoscalar form factor, $\tilde G_P(Q^2)$, enter the description of (quasi-)elastic neutrino-nucleon scattering~\cite{Ahrens:1988rr,Kitagaki:1990vs,Bodek:2007vi,Meyer:2016oeg} and exclusive pion electroproduction~\cite{Choi:1993vt,Bernard:1994pk,Liesenfeld:1999mv,Fuchs:2003vw} (e.g., $e^- p \to \pi^- p \nu$). They can also be measured in muon capture~\cite{Wright:1998gi,Winter:2011yp,Andreev:2012fj,Hill:2017wgb}. For reviews see, e.g.,  Refs.~\cite{Bernard:2001rs,Hill:2017wgb}.\par%
Apart from experimental measurements, there are various tools available that can constrain form factors from the theory side. At small virtualities chiral perturbation theory can be used to obtain valuable constraints (see, e.g., Refs.~\cite{Bernard:2001rs,Fuchs:2003vw,Schindler:2006it}). At intermediate and large~$Q^2$ the form factors can be calculated (up to some systematic uncertainty of $\mathord{\sim}15\%$) using light-cone sum rules~\cite{Braun:2006hz,Anikin:2016teg}. Alternatively, one can use functional renormalization group methods~\cite{Eichmann:2011pv}.\par
However, the cleanest method for the determination of hadron form factors is lattice QCD. Various determinations of the nucleon couplings and form factors using a wide variety of lattice actions and analysis methods can, e.g., be found in Refs.~\mbox{\cite{Martinelli:1988rr,Yamazaki:2008py,Yamazaki:2009zq,Bratt:2010jn,Alexandrou:2010hf,Capitani:2012gj,Green:2012ud,Horsley:2013ayv,Bhattacharya:2013ehc,Chambers:2014qaa,Bali:2014nma,vonHippel:2016wid,Bhattacharya:2016zcn,Meyer:2016kwb,Yoon:2016jzj,Liang:2016fgy,Bouchard:2016heu,Alexandrou:2017msl,Berkowitz:2017gql,Yao:2017fym,Chang:2018uxx,Alexandrou:2018lvq,Green:2017keo,Alexandrou:2017hac,Capitani:2017qpc,Rajan:2017lxk,Tsukamoto:2017fnm,Jang:2018lup,Ishikawa:2018rew,Liang:2018pis}}. Recent calculations that have precisely determined the axial, the pseudoscalar, and the induced pseudoscalar form factors separately from lattice data yield an unexpected result: The relation between these form factors inferred from the partial conservation of the axialvector current (denoted the \PCACFF\ relation in the following) is broken rather badly~\cite{Rajan:2017lxk,Tsukamoto:2017fnm,Ishikawa:2018rew,Jang:2018lup,Liang:2018pis}.\footnote{Note, that in Refs.~\cite{Tsukamoto:2017fnm,Ishikawa:2018rew} the PCAC relation is claimed to be satisfied, however the quark mass used differs by a factor of three from that extracted from the pion two-point functions.} Adding to the confusion, one should note that the PCAC relation itself is still fulfilled quite well on the level of the correlation functions, which leads to the conclusion that either discretization effects or excited state effects are responsible for the observed discrepancy. However, the former have been ruled out as (the sole) explanation in~\cite{Rajan:2017lxk}, while it was found in~\cite{Jang:2018lup} that even a $3$-state fit cannot resolve the issue. Let us note in passing that simply enforcing the PCAC relation on the form factor level might lead to uncontrollable systematic effects.\par%
In this work we demonstrate that the largest part of the deviation from the \PCACFF\ relation is indeed due to excited states in the temporal axialvector ($A_0$) and pseudoscalar ($P$) channels. These excited states are, however, so strongly enhanced relative to the ground state that the usual multistate fit ansatz is bound to fail for any feasible time distances between the source, the sink, and the current insertion. While this is directly visible in $A_0$ correlation functions, which are therefore usually omitted in the analysis when extracting $G_A(Q^2)$ and $\tilde G_P(Q^2)$,  up to now the problem has been overlooked in the $P$ channel (which yields $G_P(Q^2)$). However, for axialvector three-point functions, these dominant excited state contributions violate a simple relation, derived from the equation of motion, and can be removed by a straightforward projection. The PCAC relation then suggests that a similar replacement should also be implemented for pseudoscalar three-point correlation functions.\par
If the reader is now eager to learn the details of the method, he or she should skip the usual description of simulation parameters and analysis methods provided in Section~\ref{sect_sim_details}, and directly jump to Section~\ref{sect_PCACFF} where the problem and its resolution will be explained in detail. The results are then presented in Section~\ref{sect_RESULTS}, before we conclude.
\section{Simulation and analysis details}\label{sect_sim_details}
\begin{table}[t]\centering
	\caption{\label{Tab:Ensembles} Details of the ensembles used in the analysis,
		including the inverse lattice coupling $\beta$, the hopping parameter
		$\kappa$, the lattice geometry, the pion mass $m_\pi$, and the
		spatial lattice extent $L=aN_s$ in units of $m_\pi^{-1}$. The finite volume
		pion masses were determined in Ref.~\cite{Bali:2014gha} and the
		errors include an estimate of both the systematic and statistical
		uncertainty. The lattice spacings and renormalization factors are listed in Table~\ref{Tab:Renormalization}.}\vspace{0.1cm}
			\begin{widetable}{\columnwidth}{cccccc}
                \toprule
				Ens. &$\beta$&$\kappa$& $\hphantom{00^0}\mathllap{N_s^3}\times\mathrlap{N_t}\hphantom{00}$          & $m_\pi\,[\giga\electronvolt]$  &  $Lm_\pi$ \\\midrule
				I    & 5.20  & 0.13596       & $32^3\times 64$& 0.2795(18)& 3.69\\\midrule
				II   & 5.29  & 0.13620       & $24^3\times 48$& 0.4264(20)& 3.71\\
				III  &       &               & $32^3\times 64$& 0.4222(13)& 4.90\\
				IV   &       & 0.13632       & $32^3\times 64$& 0.2946(14)& 3.42\\
				V    &       &               & $40^3\times 64$& 0.2888(11)& 4.19\\
				VI   &       &               & $64^3\times 64$& 0.2895(07)& 6.71\\
				VIII &       & 0.13640       & $64^3\times 64$& 0.1497(13)& 3.47\\\midrule
				IX   & 5.40  & 0.13640       & $32^3\times 64$& 0.4897(17)& 4.81\\
				X    &       & 0.13647       & $32^3\times 64$& 0.4262(20)& 4.18\\
				XI   &       & 0.13660       & $48^3\times 64$& 0.2595(09)& 3.82\\
                \bottomrule
			\end{widetable}%
\end{table}%
\subsection{Lattice setup}
\begin{figure}[t]
	\centering\includegraphics[width=.95\columnwidth]{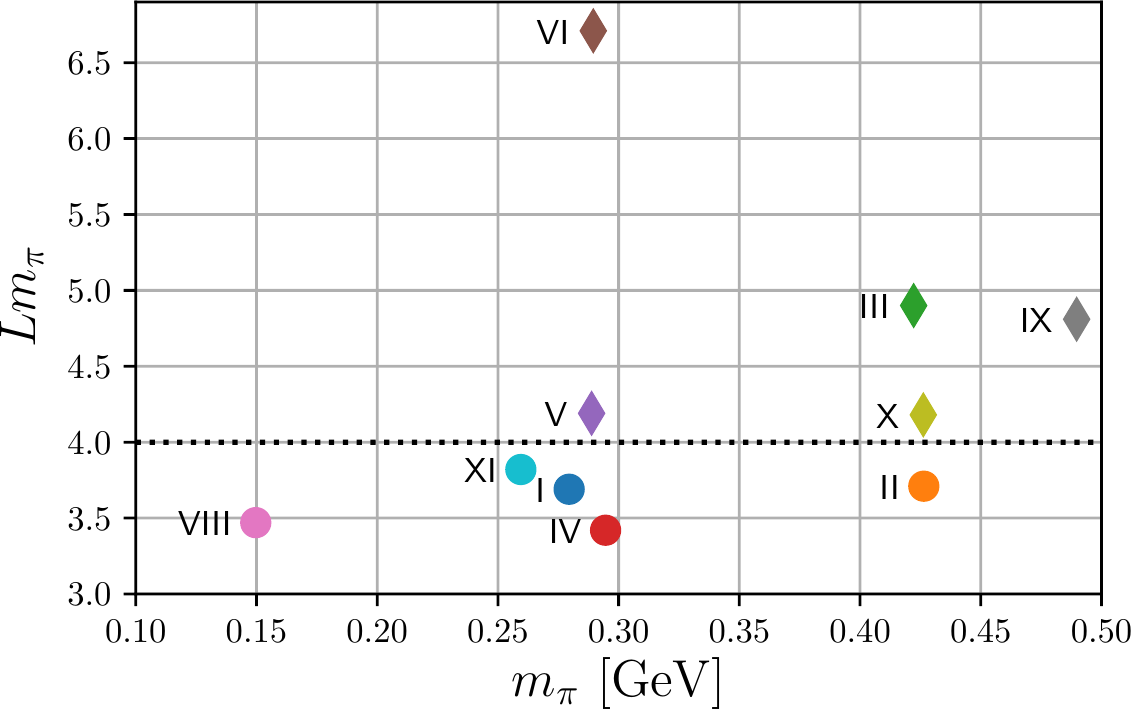}
	\caption{$Lm_\pi$ plotted against the pion mass for our ensembles listed in Table~\ref{Tab:Ensembles}. The color coding for the ensembles is used throughout this work.}
	\label{Fig:VolumeVSPion}
\end{figure}

In this work we analyse ensembles with two flavors of
nonperturbatively improved clover fermions (also known as
Sheikholeslami--Wohlert fermions~\cite{Sheikholeslami:1985ij}) and the
Wilson gauge action at three different $\beta$~values corresponding to
lattice spacings in the range of $\unit{0.060}{\femto\metre}$ to
$\unit{0.081}{\femto\metre}$. The spacing was set using the Sommer
parameter $r_0=\unit{0.5}{\femto\metre}$ determined in
Ref.~\cite{Bali:2012qs} (see also Ref.~\cite{Sommer:1993ce}). The pion
masses range from $\unit{490}{\mega\electronvolt}$ down to an almost
physical value of $\unit{150}{\mega\electronvolt}$.  The ensemble
details are provided in
Table~\ref{Tab:Ensembles} and Fig.~\ref{Fig:VolumeVSPion} visualizes
the landscape of available pion masses and volumes. The multiple
volumes for a pion mass of $\mathord{\sim}\unit{290}{\mega\electronvolt}$, ranging from $Lm_\pi=3.4$ to~$6.7$,
enable an investigation of finite volume effects. \par
The isovector form factors can be extracted from connected three-point
functions. The latter have been computed as part of a previous study
of the nucleon isovector charges~\cite{Bali:2014nma} using the
traditional sequential source method~\cite{Maiani:1987by}, where the
insertion time $t_{\rm ins}$ of the local current is varied, while the
sink and source times, $t_{\rm snk}$ and~$t_{\rm src}$,
are fixed. To increase statistics two measurements of the three-point
functions are performed at different source positions per
ensemble. Auto-correlations between configurations are taken into
account by binning with a binsize of~$10$.\par
\begin{table}[t]\centering%
			\caption{\label{Tab:Renormalization} Lattice spacings and renormalization factors~\cite{Gockeler:2010yr,Bali:2014nma}, the latter of which include the conversion to the $\MSbar$ scheme at $\mu = \unit{2}{\giga\electronvolt}$.}\vspace{0.1cm}
				\begin{widetable}{\columnwidth}{cccc}
                    \toprule
					$\beta$          &    $a\,[\femto\meter]$     &      $Z_P(\mu)$  &     $Z_A$      \\\midrule
					 5.20 & 0.081 &  0.464(12) &0.7532(16)\phantom{0}\\
					     5.29          & 0.071     & 0.476(13)     & 0.76487(64)     \\
					       5.40     & 0.060      &  0.498(09)  & 0.77756(33)   \\
					\bottomrule
				\end{widetable}
\end{table}
In order to minimize excited state contributions to the three-point
functions (and to the two-point functions that are also required),
spatially extended source and sink interpolating operators are
constructed using Wuppertal smearing~\cite{Gusken:1989qx} with
APE-smeared~\cite{Falcioni:1984ei} gauge links. In
Ref.~\cite{Bali:2014nma} the smearing was optimized such that ground
state dominance was observed in the nucleon two-point functions at
around the same physical time, $t=t_{\rm snk}-t_{\rm src}\approx\unit{0.8}{\femto\meter}$, for
different pion masses and lattice spacings. For key ensembles,
labelled~III, IV, and~VIII, corresponding to pion masses of~$420$, $290$,
and~$\unit{150}{\mega\electronvolt}$, respectively, at the lattice spacing $a=\unit{0.071}{\femto\meter}$, multiple
source-sink separations for the three-point functions were generated
to enable an investigation of remaining excited state contamination.
The separations correspond to $t\approx\unit{(1.1,1.2)}{\femto\meter}$ for ensemble~III,
$t\approx\unit{(0.5,0.6,0.8,0.9,1.1,1.2)}{\femto\meter}$ for ensemble~IV and $t\approx\unit{(0.6,0.9,1.1)}{\femto\meter}$
for ensemble~VIII. The analysis of the isovector charges indicated
that, for the smearing applied, the ground state contribution could be
reliably determined for $t\gtrsim\unit{1}{\femto\meter}$ and this (single) source-sink
separation was employed for the remaining ensembles.  \par
The simultaneous fits to the two- and three-point functions for the
present analysis, taking into account the leading excited state
contribution, are discussed in Section~\ref{Sect:ExcitedStates}.\par%
\subsection{Correlation functions and form factor decompositions}
        We analyse the two- and three-point functions
        \begin{align}
            C_{\rm 2pt}^{\vec{p}} (t) &= a^3 \sum_{\vec{x}} e^{-i\vec{p}\vec{x}} P_+^{\alpha\beta}\la O^\beta_N(\vec{x}, t) \bar{O}^{\alpha}_N(\vec{0}, 0) \ra \,, \label{Eqs:TwopointFunctionDefinition} \\
            C_{{\rm 3pt}, \Gamma}^{\vec{p}^{\mathrlap{\prime}}, \vec{p}, \mathcal{O}} (t, \tau) &= a^6 \smashoperator{\sum_{\vec{x}, \vec{y}}} e^{-i\vec{p}^{\prime \!}\vec{x} + i(\vec{p}^{\mathrlap{\prime}} - \vec{p})\vec{y}} \notag \\
			   &\quad\times\Gamma^{\alpha\beta} \la O^\beta_N(\vec{x}, t) \mathcal{O}(\vec{y}, \tau) \bar{O}^{\alpha}_N(\vec{0}, 0) \ra \,, \label{Eqs:ThreepointFunctionDefinition}
        \end{align}
        where $t=t_{\rm snk}-t_{\rm src}$, $\tau=t_{\rm ins}-t_{\rm src}$, $O_N=(u^TC\gamma_5 d) u$ is the Wuppertal-smeared interpolating current for the nucleon with the charge conjugation matrix $C$, and $P_+ = (1 + \gamma_0)/2$ projects onto positive parity for zero momentum. We choose $\Gamma$ to be $P_+^i = P_+^{\phantom{i}} \gamma^i \gamma_5$, $i=1,2,3$, in our analysis. In our actual simulations, we restrict the kinematics to $\vec{p}^\prime=\vec{0}$.\par
        Inserting complete sets of states, the correlation functions in Euclidean time can be expanded in terms of had\-ronic matrix elements. For the two-point function this yields
		\begin{align}
			\begin{split}
				C_{\rm 2pt}^{\vec{p}} (t) &= \sum\limits_\sigma P_+^{\alpha\beta} \la 0 | O_N^\beta | N_\sigma^{\vec{p}} \ra \la N_\sigma^{\vec{p}} | \bar{O}_N^\alpha |  0 \ra \frac{e^{-E_{\vec{p}}t}}{2E_{\vec{p}}} + \dots \\
				&= Z_{\vec{p}}  \frac{E_{\vec{p}} + m_N}{E_{\vec{p}}} e^{-E_{\vec{p}}t} + \dots \,,
			\end{split} \label{Eqs:TwopointFunctionGroundState}
		\end{align}
		where only the ground state contribution is given and $m_N$ denotes the nucleon mass. The excited state corrections are discussed in Sect.~\ref{Sect:ExcitedStates}. The normalization $Z_{\vec{p}}$ is smearing-dependent and encodes the overlap of the ground state with the interpolating operators at the source and the sink,
		\begin{equation}
            \la 0 | O_N^\beta | N_\sigma^{\vec{p}} \ra = \sqrt{Z_{\vec p}} \, u^\beta_{\vec{p}, \sigma} \,,
		\end{equation}
		where $u^\beta_{\vec{p}, \sigma}$ is a nucleon spinor.\par
		An analogous spectral decomposition of the three-point functions with an operator insertion $\mathcal O \in \{P,A_\mu\}$ corresponding to the isovector pseudoscalar and axialvector currents,
		\begin{align}
		P &= \bar u \gamma_5 u - \bar d \gamma_5 d \,, &
		A_\mu &= \bar u \gamma_\mu \gamma_5 u -  \bar d \gamma_\mu \gamma_5 d \,,
		\end{align}
		leads to
		\begin{align}
			C_{{\rm 3pt}, \Gamma}^{\vec{p}^{\mathrlap{\prime}}, \vec{p}, \mathcal{O}} (t, \tau) &=
			\sqrt{ Z_{\vec{p}^\prime} Z_{\vec{p}} } \, B_{\Gamma,\mathcal{O}}^{\vec{p}^{\mathrlap{\prime}}, \vec{p}} \, e^{-E_{\vec{p}^\prime} (t-\tau)}e^{- E_{\vec{p}} \tau} + \dots \label{Eqs:ThreepointFunctionGroundState}
		\end{align}
        with
		\begin{align}
			 B_{\Gamma,\mathcal{O}}^{\vec{p}^{\mathrlap{\prime}}, \vec{p}}& = \frac{1}{4E_{\vec{p}^\prime}E_{\vec{p}}} \operatorname{Tr} \bigl\{ \Gamma (\slashed{p}^\prime+m_N) J[\mathcal{O}] (\slashed{p}+m_N) \bigr\} \,.\label{Eqs:Bamp}
		\end{align}
		$J[\mathcal{O}]$ is defined by the form factor decomposition
		\begin{align}
		\la N^{\smash{\vec p^\prime}}_{\smash{\sigma^\prime}} | \mathcal O | N^{\vec p}_{\sigma} \ra &= \bar u_{\vec{p}^{\mathrlap{\prime}}, \sigma^\prime} J[\mathcal{O}]  u_{\vec{p}, \sigma} \,.
		\end{align}
		For the different channels it reads%
		\begin{align}%
			J[P] &= \gamma_5 G_P(Q^2) \,, \label{Eqs:PseudoscalarFF}\\
			J[A_\mu] &= \gamma_\mu \gamma_5 G_A(Q^2) + \frac{q_\mu}{2m_N} \gamma_5 \tilde{G}_P(Q^2) \,, \label{Eqs:AxialFF}
		\end{align}%
where $q=p^\prime-p$ and $Q^2 = -q^2$.%
\par%
\pagebreak[2]%
		The axial Ward identity yields a partial conservation of the axialvector current, $\partial^\mu\!A_\mu = 2i \, m_q P$, known as the PCAC relation. On the lattice this relation can be broken by discretization effects. For the nucleon matrix elements it implies:
		\begin{equation}
			2i \, m_{q} \la N^{\smash{\vec p^\prime}}_{\smash{\sigma^\prime}} | P | N_\sigma^{\vec{p}} \ra = \la N^{\smash{\vec p^\prime}}_{\smash{\sigma^\prime}} | \partial^\mu\!A_\mu | N_\sigma^{\vec{p}} \ra +\mathcal{O}(a^2) \,.\label{Eqs:NucleonMatrixPCAC}
		\end{equation}
		Using the definitions~\eqref{Eqs:PseudoscalarFF} and~\eqref{Eqs:AxialFF} together with the equations of motion one can deduce the corresponding relation for the form factors (\PCACFF):
		\begin{equation}
			m_q G_P(Q^2) = m_N G_A(Q^2) - \frac{Q^2}{4m_N} \tilde{G}_P(Q^2) + \mathcal{O}(a^2) \,.  \label{Eqs:NucleonFFPCAC}
		\end{equation}
		Eqs.~\eqref{Eqs:NucleonMatrixPCAC} and~\eqref{Eqs:NucleonFFPCAC} should be satisfied to a similar degree once the ground state matrix elements have been extracted reliably.%
	\subsection{Excited states analysis}\label{Sect:ExcitedStates}
	    In three-point functions the signal-to-noise ratio decreases exponentially with the time distance between the source and the sink. Hence, within typical separations $t\lesssim \unit{1.5}{\femto\meter}$ a sufficient suppression of excited states may not be achieved. Including the leading excited state contributions to Eqs.~\eqref{Eqs:TwopointFunctionGroundState} and~\eqref{Eqs:ThreepointFunctionGroundState} gives:
		\begin{figure}[tb]%
			\centering\includegraphics[width=.9\columnwidth]{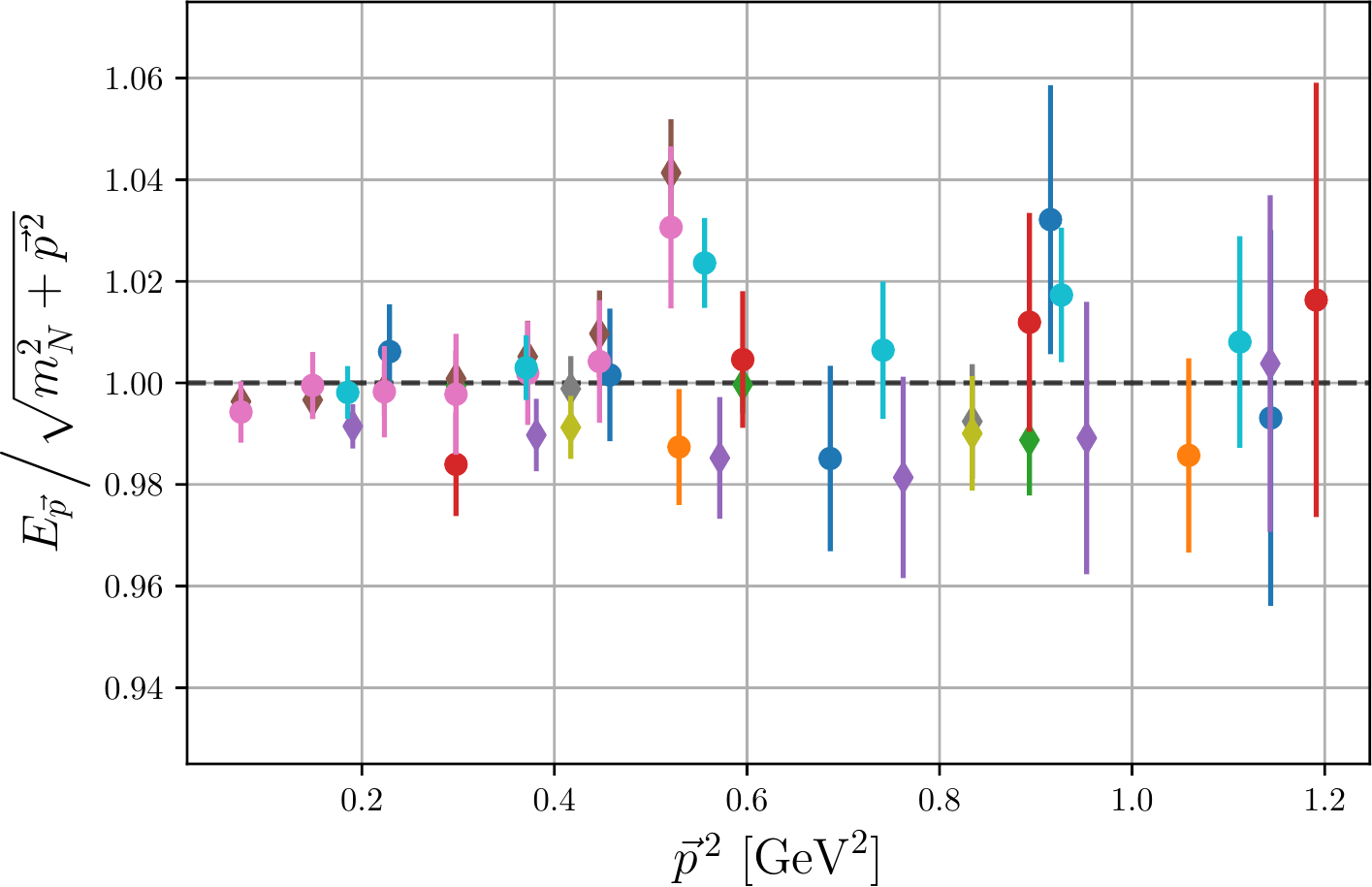}%
			\caption{Lattice data of the ground state nucleon energy, normalized to the continuum expectation~\eqref{Eqs:DispersionRel}. The color coding follows Fig.~\ref{Fig:VolumeVSPion}.\label{Fig:DispersionRel}}%
		\end{figure}%
		\begin{align}
			C_{\rm 2pt}^{\,\vec{p}} (t) &=  Z_{\vec{p}}  \frac{E_{\vec{p}} + m_N}{E_{\vec{p}}} e^{-E_{\vec{p}}t} \Bigl( 1 + \tilde{Z}  e^{-\Delta E_{\vec{p}}t} \Bigr)\,, \label{Eqs:FitFunctionC2pt}\\
			\begin{split}
				C_{{\rm 3pt}, \Gamma}^{\vec{p}^{\mathrlap{\prime}}, \vec{p}, \mathcal{O}} (t, \tau) &= \sqrt{ Z_{\vec{p}^\prime} Z_{\vec{p}} } \, B_{\Gamma,\mathcal{O}}^{\vec{p}^{\mathrlap{\prime}}, \vec{p}} \, e^{-\Ef(t-\tau)} e^{-E_{\vec{p}}\tau} \\
				&\quad \times \Bigl( \begin{aligned}[t]  1 &+ B_{10} e^{-\Delta\Ef(t-\tau)}+ B_{01} e^{-\Delta E_{\vec{p}}\tau}  \\
     		    &+ B_{11} e^{-\Delta\Ef(t-\tau)} e^{-\Delta E_{\vec{p}}\tau} \Bigr) \,, \label{Eqs:FitFunctionC3pt} \end{aligned}
     		\end{split}
		\end{align}
		\begin{figure*}[t]\centering
		\begin{minipage}[t]{0.64\textwidth}%
			\centering\includegraphics[clip,trim=0 20 0 0,width=\textwidth]{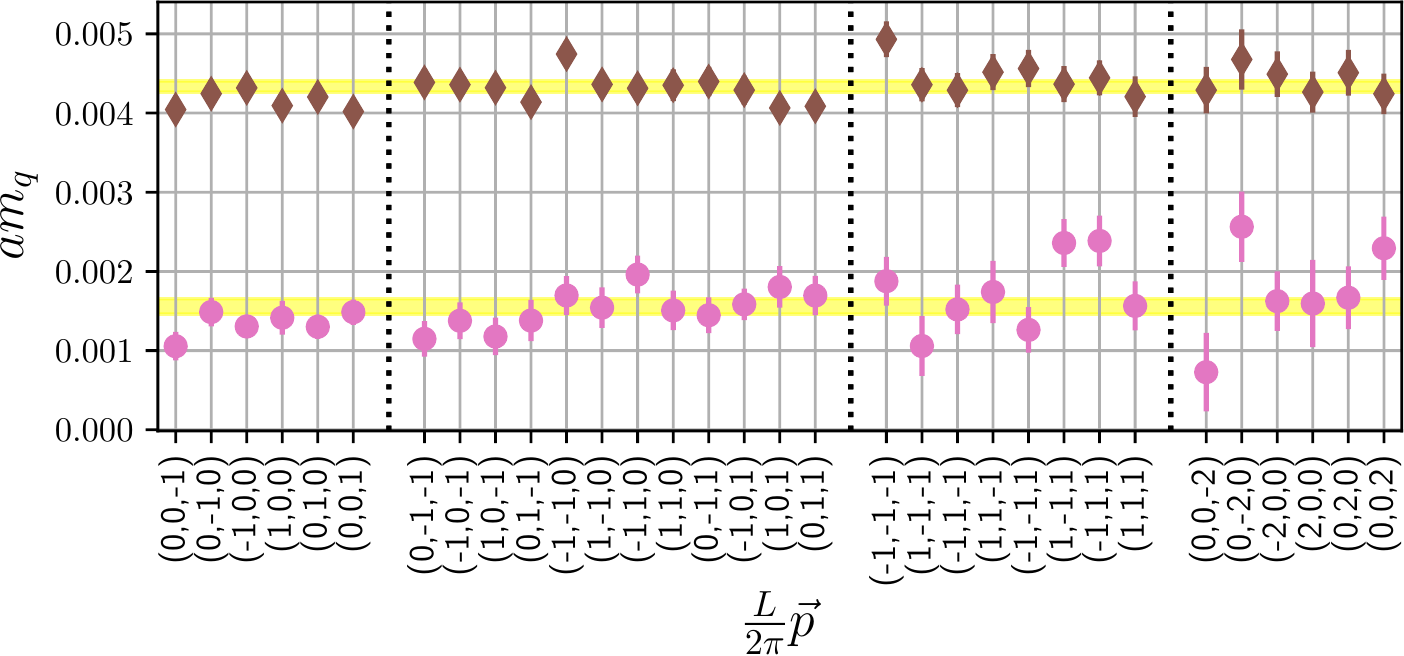}%
		\end{minipage}\hfill\begin{minipage}[b]{0.32\textwidth}%
			\caption{\label{Fig:PCACMassEns0608}Quark mass obtained from the ratio (see Eq.~\eqref{Eqs:PCACMassFromNucleon3pts}) utilizing the PCAC relation~\eqref{Eqs:PCACRelation} for the ensembles VI (diamonds) and VIII (circles). The result is given for various initial momenta $\vec p$ in units of $2\pi/L$, while the final momentum is always fixed to $\vec p^\prime=\vec 0$ in our kinematics. Indeed, the PCAC relation is valid on the three-point function level, up to $\mathcal{O}(a^2)$ effects.}\vspace{1.25cm}\end{minipage}
		\end{figure*}%
                where $\Delta E_{\vec{p}}$ denotes the energy gap between the first excited state and the ground state. The excited state coefficient $\tilde{Z}$ depends on the nucleon interpolator, its smearing, and the momenta, while $B_{10}$, $B_{01}$, and $B_{11}$ also depend on the current $\mathcal{O}$ and on the projector $\Gamma$.\par
		For illustrative purposes we define the ratio
		\begin{align}
			R_{\Gamma, \mathcal{O}}^{\vec{p}^{\mathrlap{\prime}}, \vec{p}} (t, \tau) &=
			\frac{ C_{{\rm 3pt}, \Gamma}^{\vec{p}^{\mathrlap{\prime}}, \vec{p}, \mathcal{O}} (t, \tau) }{ C_{\rm 2pt}^{\vec{p}'} (t) }
			\sqrt{\frac{
				C_{\rm 2pt}^{\vec{p}'} (\tau) C_{\rm 2pt}^{\vec{p}'} (t) C_{\rm 2pt}^{\vec{p} } (t-\tau)
			}{
				C_{\rm 2pt}^{\vec{p} } (\tau) C_{\rm 2pt}^{\vec{p} } (t) C_{\rm 2pt}^{\vec{p}'} (t-\tau)
			}} \nonumber\\ \label{ratio}
			&\overset{t \gg \tau \gg 0}{\longrightarrow}  \sqrt{\frac{E_{\vec{p}^\prime}E_{\vec{p}} }{ (E_{\vec{p}'} + m_N)(E_{\vec{p}} + m_N)}} B_{\Gamma,\mathcal{O}}^{\vec{p}^{\mathrlap{\prime}}, \vec{p}} \,,
		\end{align}
		which eliminates the leading order time dependence as well as the overlap factors.
		Our ground state energies are well described by the continuum dispersion relation
		\begin{equation}
			E_{\vec{p}} = \sqrt{\smash{m_N^2 + \vec{p}^{\,2}}\rule[-1pt]{0sp}{10pt}} \,, \label{Eqs:DispersionRel}
		\end{equation}%
		as shown in Fig.~\ref{Fig:DispersionRel} and we assume this functional form in our fitting analysis. Excited states will in general contain more than one hadron and hence we make no such assumption for $\Delta E_{\vec{p}}$. We remark that the spectrum includes $N\pi$, $N\pi\pi$, and higher states and that this spectrum becomes more dense as the pion mass decreases. At zero momentum the lowest multiparticle excited states are $P$\nobreakdash-wave~$N\pi$ and $S$\nobreakdash-wave~$N\pi\pi$.\par
For a given momentum transfer $q^2$, a simultaneous fit of the form of Eqs.~\eqref{Eqs:FitFunctionC2pt} and~\eqref{Eqs:FitFunctionC3pt} is performed to the relevant two- and three-point functions, including all available momentum directions, hadron polarizations, and source-sink separations. The form factors enter the fit directly as parameters, substituting the amplitudes $B_{\Gamma,\mathcal{O}}^{\vec{p}^{\mathrlap{\prime}}, \vec{p}}$ utilising Eqs.~\mbox{\eqref{Eqs:Bamp}--\eqref{Eqs:AxialFF}} and the dispersion relation. For ensembles with only one value of $t$, the parameter $B_{11}$ is set to zero. The fit range is chosen to be $2a\leq \tau\leq t-2a$ resulting in reasonable values of $\chi^2/\text{d.o.f}$.
	\subsection{Renormalization and \texorpdfstring{$\mathcal O(a)$}{order a} improvement}
		The renormalization factors have been calculated nonperturbatively in an RI${}^\prime$-MOM scheme using the Rome--Southampton method~\cite{Martinelli:1994ty} and then converted into the $\MS$ scheme using three-loop continuum perturbation theory. A detailed discussion can be found in~\cite{Gockeler:2010yr}.\par
		The isovector currents are multiplicatively renormalized using
		\begin{equation}
			\mathcal{O}^{\text{ren}}_X = Z_X^{}(\beta)\Big[1+am_q b_X^{}(\beta)\Big] \mathcal{O}_X^{\smash{\text{imp}}}\,,
		\end{equation}
		where the relevant $Z_X$, listed in Table \ref{Tab:Renormalization}, contain both the nonperturbative renormalization and the conversion to the $\MS$ scheme at a scale of $\mu = \unit{2}{\giga\electronvolt}$. The one-loop improvement coefficients $b_X$ have been calculated perturbatively in Refs.~\cite{Sint:1997jx,Taniguchi:1998pf} and are close to unity. The numeric values for our lattices are provided in Ref.~\cite{Bali:2014nma}. Within the Symanzik improvement program~\cite{Symanzik:1983dc,Symanzik:1983gh} also the currents themselves have to be $\mathcal O(a)$-improved. For the axialvector current this yields
		\begin{equation}
			A_\mu^{\smash{\text{imp}}} = A_\mu^{} + c_A a\partial_\mu P \,,
		\end{equation}
		where we used the improvement coefficient $c_A$, nonperturbatively determined in Ref.~\cite{DellaMorte:2005aqe}, and $\partial_\mu$ denotes the symmetrically discretized derivative. Note that for the pseudoscalar current $P^{\text{imp}} = P$.\par

\section{PCAC on the form factor level}\label{sect_PCACFF}
	\subsection{Quark mass from nucleon correlators}\label{sect_quark_mass}
	Since the partial conservation of the axialvector current,
        \begin{equation} \label{Eqs:PCACRelation}
	  \partial^\mu\!A_\mu = 2i \, m_{q} P\,,
	\end{equation}
	is an operator relation, it has to hold on the correlation function level such that, using the three-point functions defined by Eq.~\eqref{Eqs:ThreepointFunctionDefinition}, the PCAC quark mass can be determined as
	\begin{equation} \label{Eqs:PCACMassFromNucleon3pts}
	 m_q = \frac{\partial^\mu  C_{{\rm 3pt}, \Gamma}^{\vec{p}^{\mathrlap{\prime}}, \vec{p}, A_\mu}  }{2i\,  C_{{\rm 3pt}, \Gamma}^{\vec{p}^{\mathrlap{\prime}}, \vec{p}, P} } \,,
	\end{equation}
	independent of any spectral analysis. As no significant $\mathcal{O}(a^2Q^2)$ effects are observed within the statistical error, we determine the quark mass by fitting to several $Q^2$ simultaneously (see Fig.~\ref{Fig:PCACMassEns0608}). Note that the spatial derivatives can be calculated using the formula
	\begin{align} \label{derivative_spatial}
	   \la \vec p^\prime | \partial^i \mathcal O | \vec p \ra &= -i \frac{\sin(aq^i)}{a} \la \vec p^\prime | \mathcal O | \vec p \ra \,,
	\end{align}
	as long as one considers only states $| \vec p \ra$ for which the external three-momenta have been fixed via an appropriate Fourier transform in the correlation function~\eqref{Eqs:ThreepointFunctionDefinition}. The time derivative has to be calculated explicitly, since the energy in the correlation function is not fixed.\par
	Usually the PCAC mass is obtained from ratios of pion two-point functions, where one can achieve very small statistical errors. Up to discretization effects, one would expect these values to agree with those obtained from ratios of baryonic three-point functions via Eq.~\eqref{Eqs:PCACMassFromNucleon3pts}. This is indeed the case, as depicted in Fig.~\ref{Fig:PCACMassExtrapolation} using ensembles I, IV, and XI (which have different lattice spacings, but very similar pion masses and volumes). Note that despite the unphysical pion mass the extrapolated ratio compares reasonably well with the $N_f=2$ value in the physical limit: $m_q/m_{\pi}^2=\unit{0.198(11)}{\giga\electronvolt^{-1}}$~\cite{Aoki:2016frl}.\par
	\subsection{Uncovering the ground state contribution}\label{sect_uncoverGS}
\begin{figure}[t]%
			\centering\includegraphics[width=.95\columnwidth]{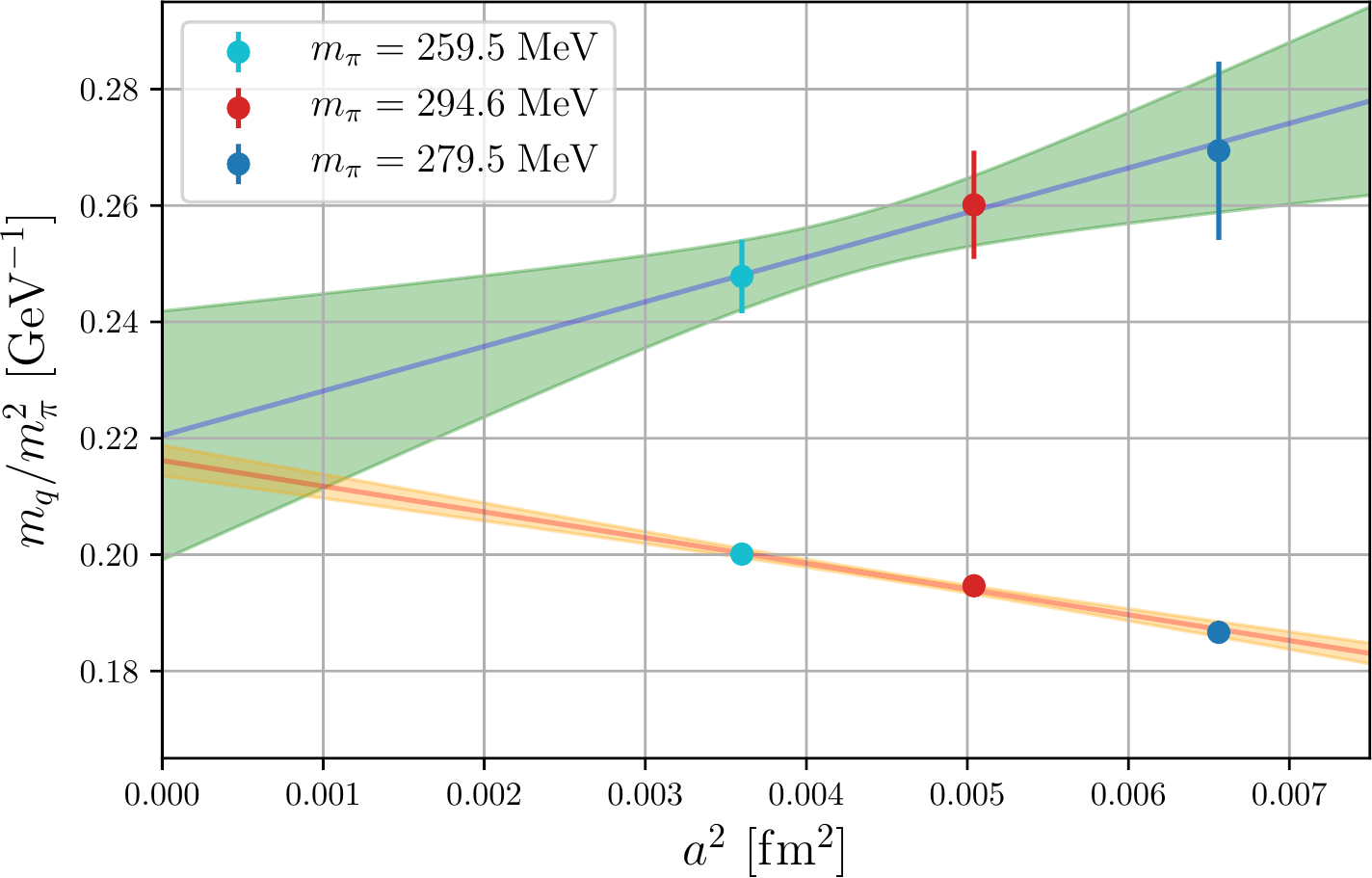}%
			\caption{\label{Fig:PCACMassExtrapolation}The continuum extrapolations of the PCAC mass determined from pion two-point functions (orange band) and nucleon three-point functions (green band) are consistent within the statistical errors. Both determinations show the leading quadratic behaviour characteristic for $\mathcal{O}(a)$-improved currents. Dividing the quark mass by the pion mass squared accounts for the leading pion mass dependence given by the Gell-Mann--Oakes--Renner relation.}%
		\end{figure}

	In a number of lattice simulations it has been observed that, even though the PCAC relation is fulfilled  on the correlation function level up to discretization effects (in our case of $\mathcal{O}(a^2)$, cf.\ Figs.~\ref{Fig:PCACMassEns0608} and \ref{Fig:PCACMassExtrapolation}), the equivalent equation for the nucleon form factors, Eq.~\eqref{Eqs:NucleonFFPCAC}, seems to be broken rather badly. Note that this problem cannot be solved just by using the PCAC mass obtained from the ratio of nucleon three-point functions discussed above.\par
\begin{figure}[tb]\centering%
 		\includegraphics[width=.95\columnwidth]{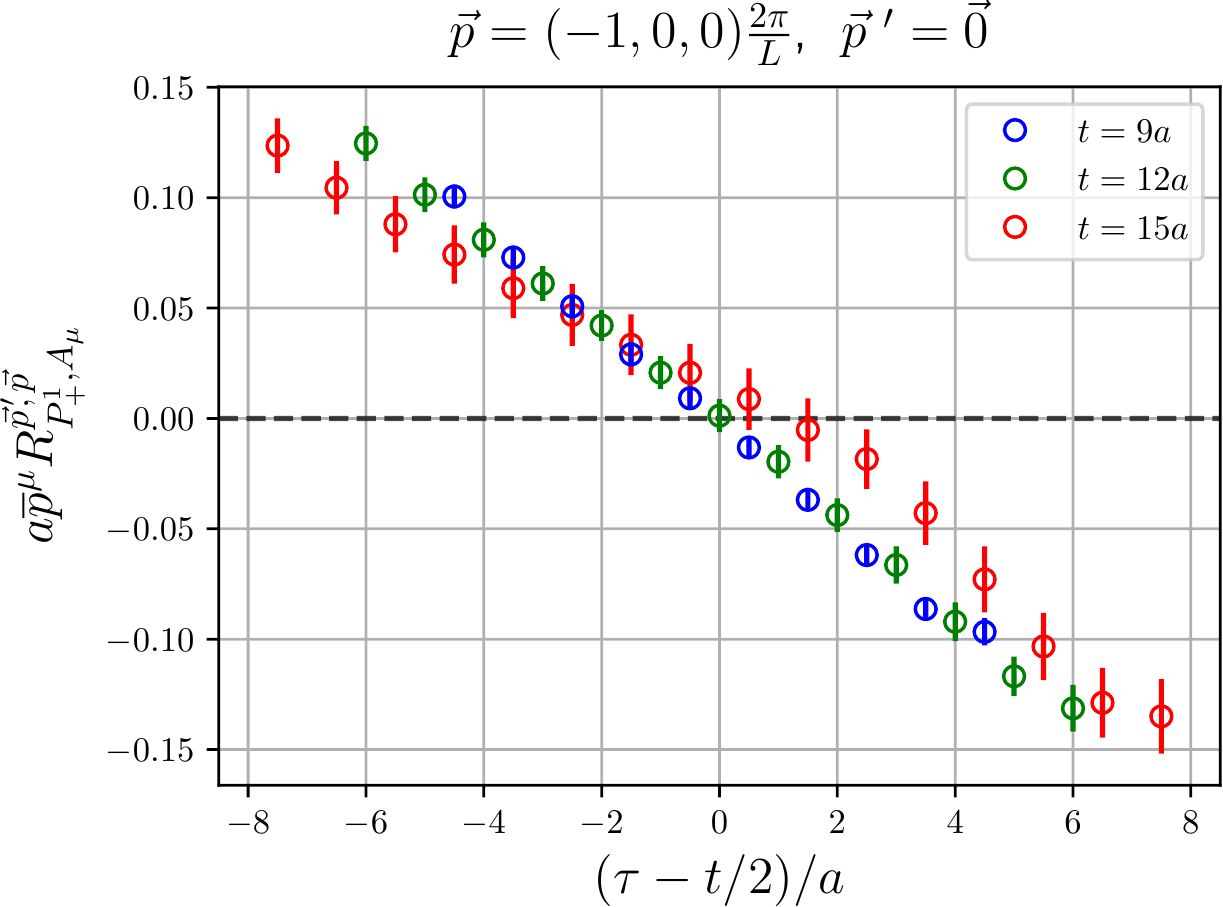}
 		\caption{Ratio of correlation functions (cf.\ Eq.~\eqref{ratio}) corresponding to the l.h.s.\ of Eq.~\eqref{PmuAmu} on ensemble~VIII, showing the dominant excited state effects.\label{PlotPmuAmu}}
 	\end{figure}

\begin{figure*}[tb]
			\centering\includegraphics[height=0.157\textheight]{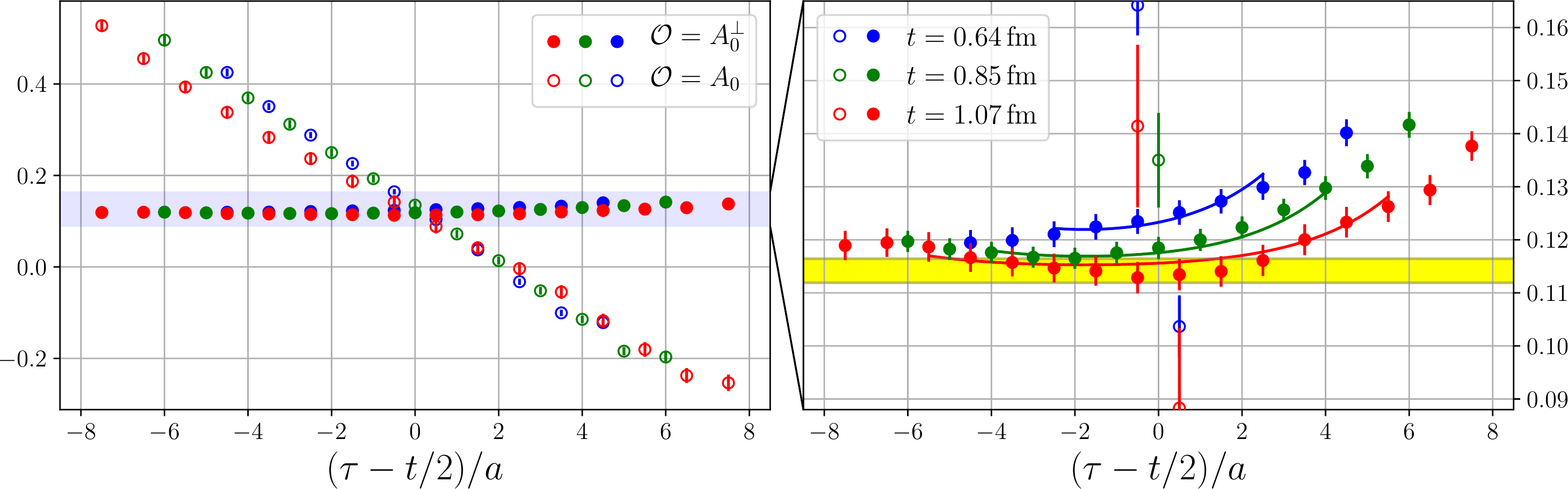}\hfill\includegraphics[height=0.157\textheight]{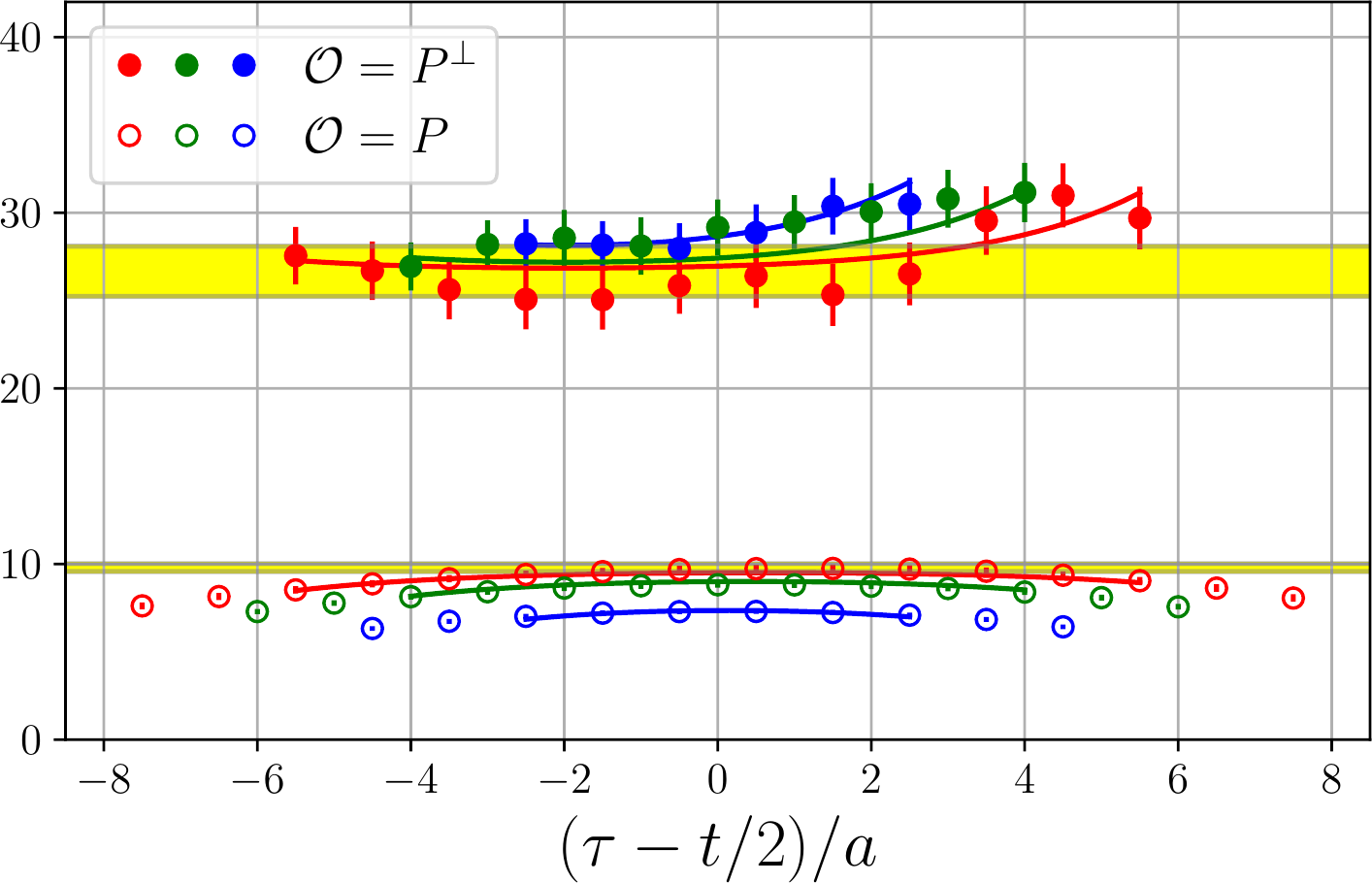}%
			\caption{Correlation function ratios (cf.\ Eq.~\eqref{ratio}) for the axialvector (left, middle) and pseudoscalar currents (right) with three different source sink separations \mbox{$t/a = 9\text{ (blue), } 12\text{ (green), } 15\text{ (red)}$} for ensemble VIII with $m_\pi \approx \unit{150}{\mega\electronvolt}$. In both cases we compare results from the standard current (open symbols) and our modified current (filled symbols). The shown results correspond to an average over all relevant combinations of polarizations and momenta with $|\vec p|=2\pi/L$ and $\vec p^\prime=\vec 0$ (i.e., $Q^2=\unit{0.073}{\giga\electronvolt\squared}$). For $A_0$ the problem is clearly visible. The signal for $A_0^\perp$ has a significantly smaller statistical error and only shows mild excited state contributions (middle, zoomed), which are resolvable with the multiexponential ansatz given in Eqs.~\eqref{Eqs:FitFunctionC2pt} and~\eqref{Eqs:FitFunctionC3pt}. In contrast, the extent of the excited state contaminations to the data for the pseudoscalar current $P$ is not so obvious. However, subtracting the same excited states causing the problem in the axialvector channel (by using $P^\perp$), one finds that the true ground state plateau lies much higher. The yellow bands indicate the ground state contributions extracted from the fits for $\mathcal{O} \in \{A_0^\perp, P, P^\perp\}$.\label{CompareG4G5}}
	\end{figure*}

%
	In the following we will demonstrate that the breaking of the PCAC relation on the form factor level is a consequence of very large excited state effects that cannot be resolved by the standard excited state analysis described in Section~\ref{Sect:ExcitedStates}. Using Eq.~\eqref{Eqs:AxialFF} together with $ (\slashed{p} - m_N) u_{\vec p,\sigma} = 0$
	one finds for the nucleon ground state:
	\begin{align}
		\pbar^\mu \la N^{\smash{\vec p^\prime}}_{\smash{\sigma^\prime}} | A_\mu | N^{\vec p}_{\sigma}  \ra = 0 \,,  \label{PmuAmu}
	\end{align}
	where $\pbar = \frac12 (p^\prime +  p)$. From Fig.~\ref{PlotPmuAmu} one can easily verify that this equation is violated by the data, demonstrating the presence of large excited state contaminations. This is mostly due to the $A_0$ correlation function, which has an almost linear dependence on the insertion time, cf.\ the left panel of Fig.~\ref{CompareG4G5}. \par
\begin{figure}[tb]\centering%
		\includegraphics[width=\columnwidth]{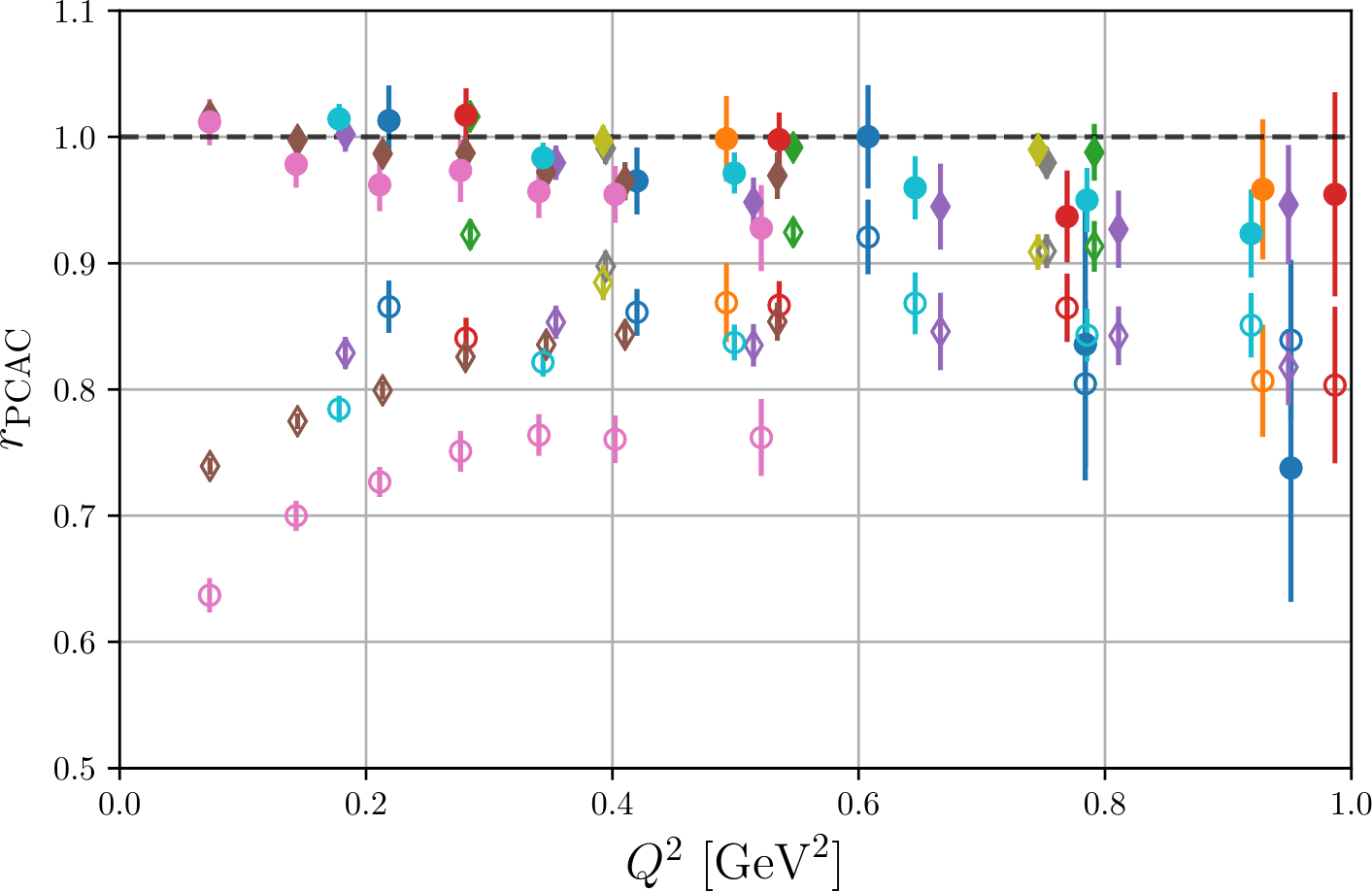}%
		\caption{\label{PlotRatiosPCAC}Violation of \PCACFF. The plot shows the ratio defined in Eq.~\eqref{eq:ratioPCAC}. The filled (open) points are obtained with (without) the excited state subtraction described in Sect.~\ref{sect_uncoverGS}. The color coding follows Fig.~\ref{Fig:VolumeVSPion}.}
	\end{figure}
\begin{figure}[tb]\centering%
		\includegraphics[width=\columnwidth]{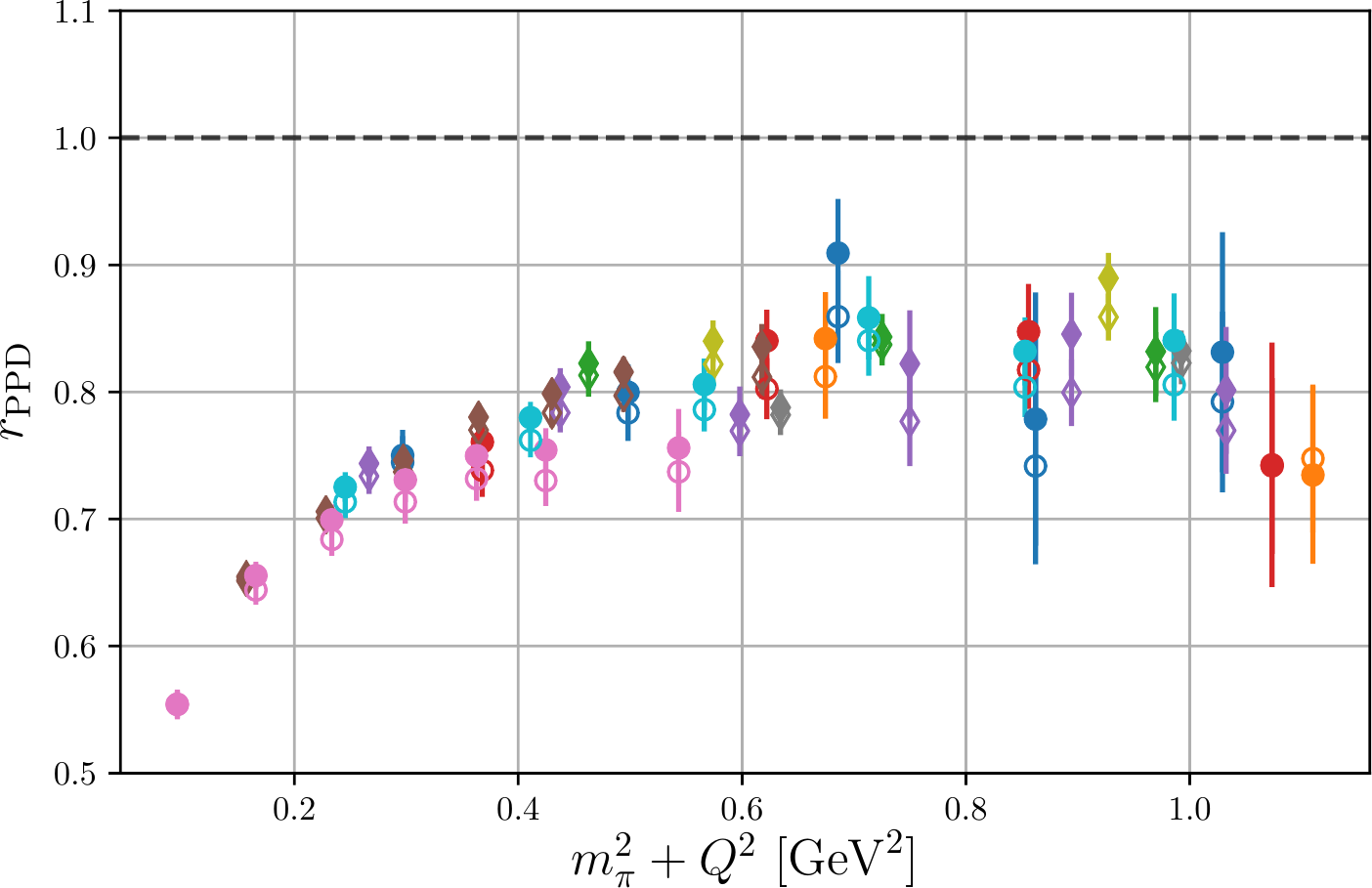}%
		\caption{\label{PlotRatiosPPD}Violation of the pion pole dominance ansatz~\eqref{Eq:PPD} for the induced pseudoscalar form factor. The filled and open data points are obtained with and without the excited state subtraction described in Sect.~\ref{sect_uncoverGS}, respectively, and agree within errors. The color coding follows Fig.~\ref{Fig:VolumeVSPion}.}%
	\end{figure}
\begin{figure*}[tb]\centering
{\includegraphics[width=0.325\linewidth]{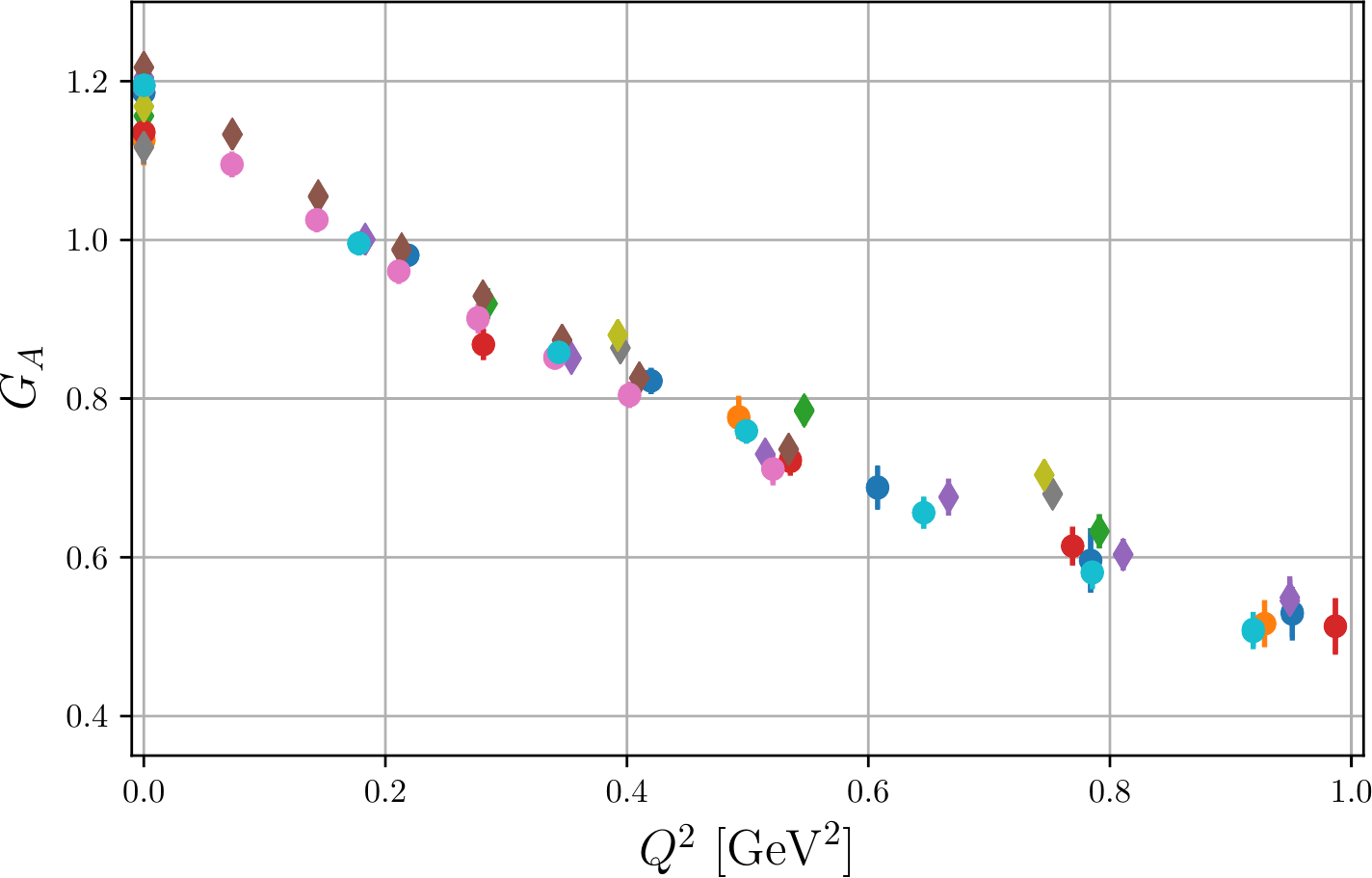}}\hfill
{\includegraphics[width=0.325\linewidth]{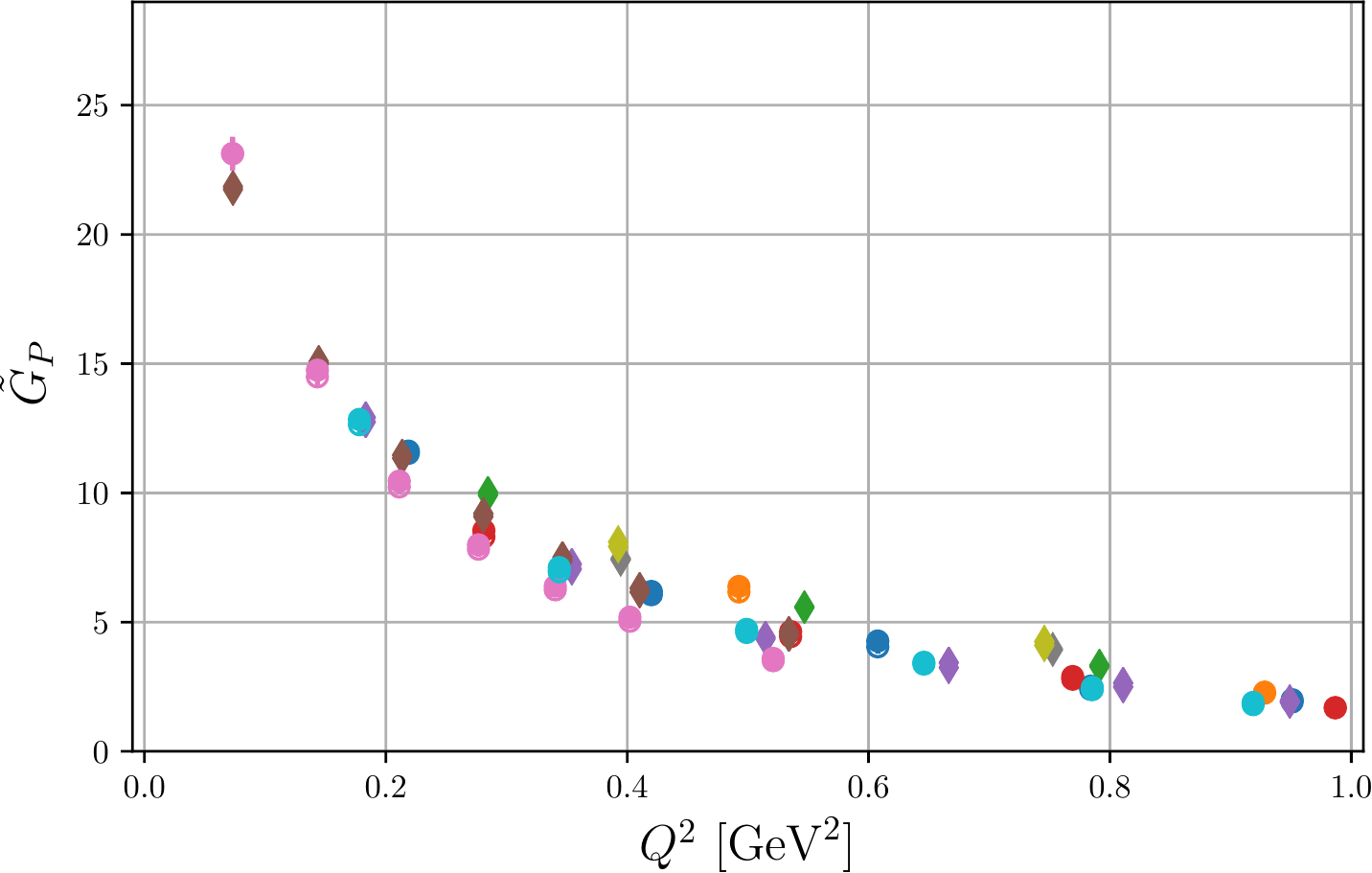}}\hfill
{\includegraphics[width=0.325\linewidth]{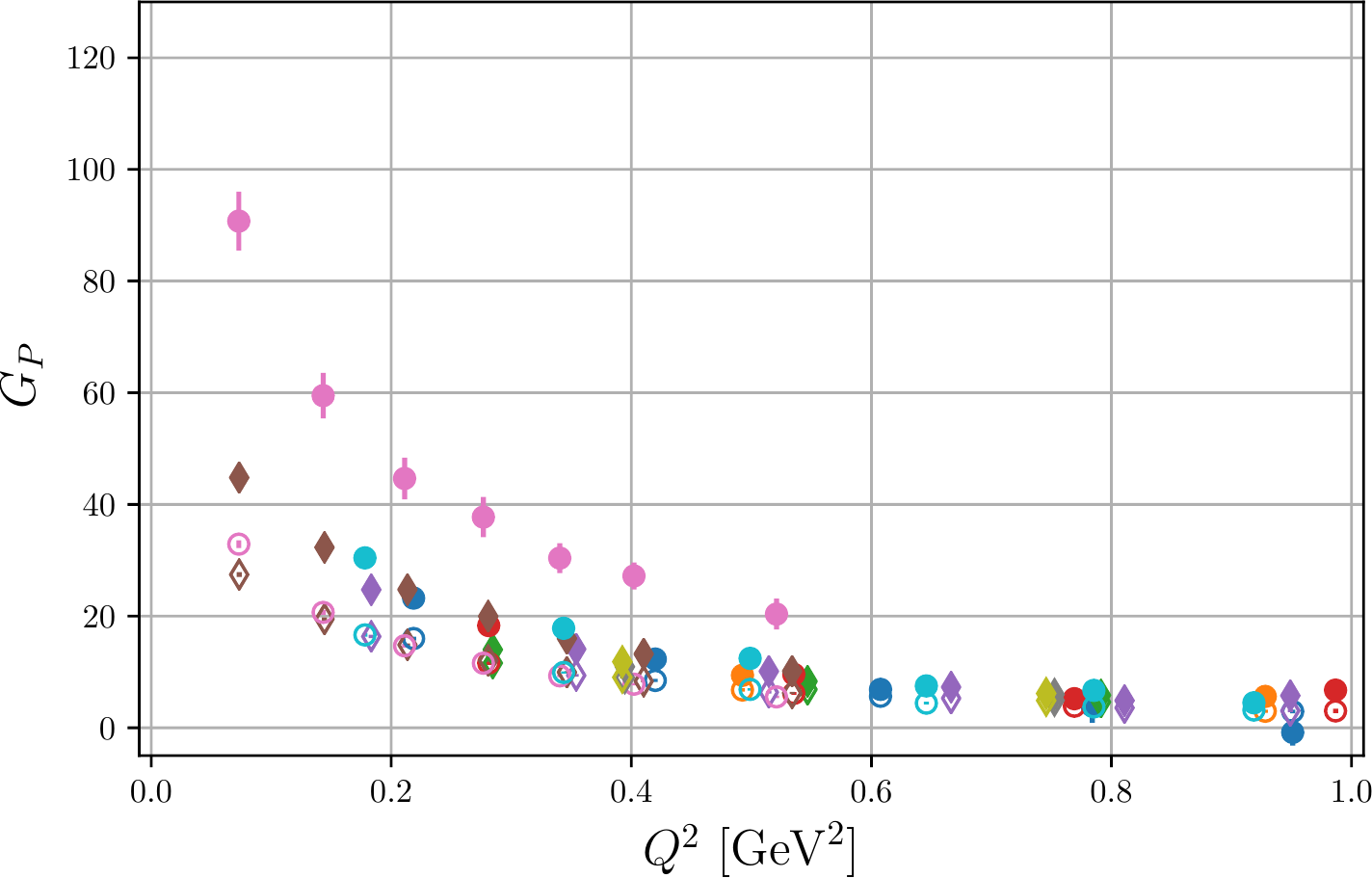}}\\[0.1cm]
{\includegraphics[width=0.325\linewidth]{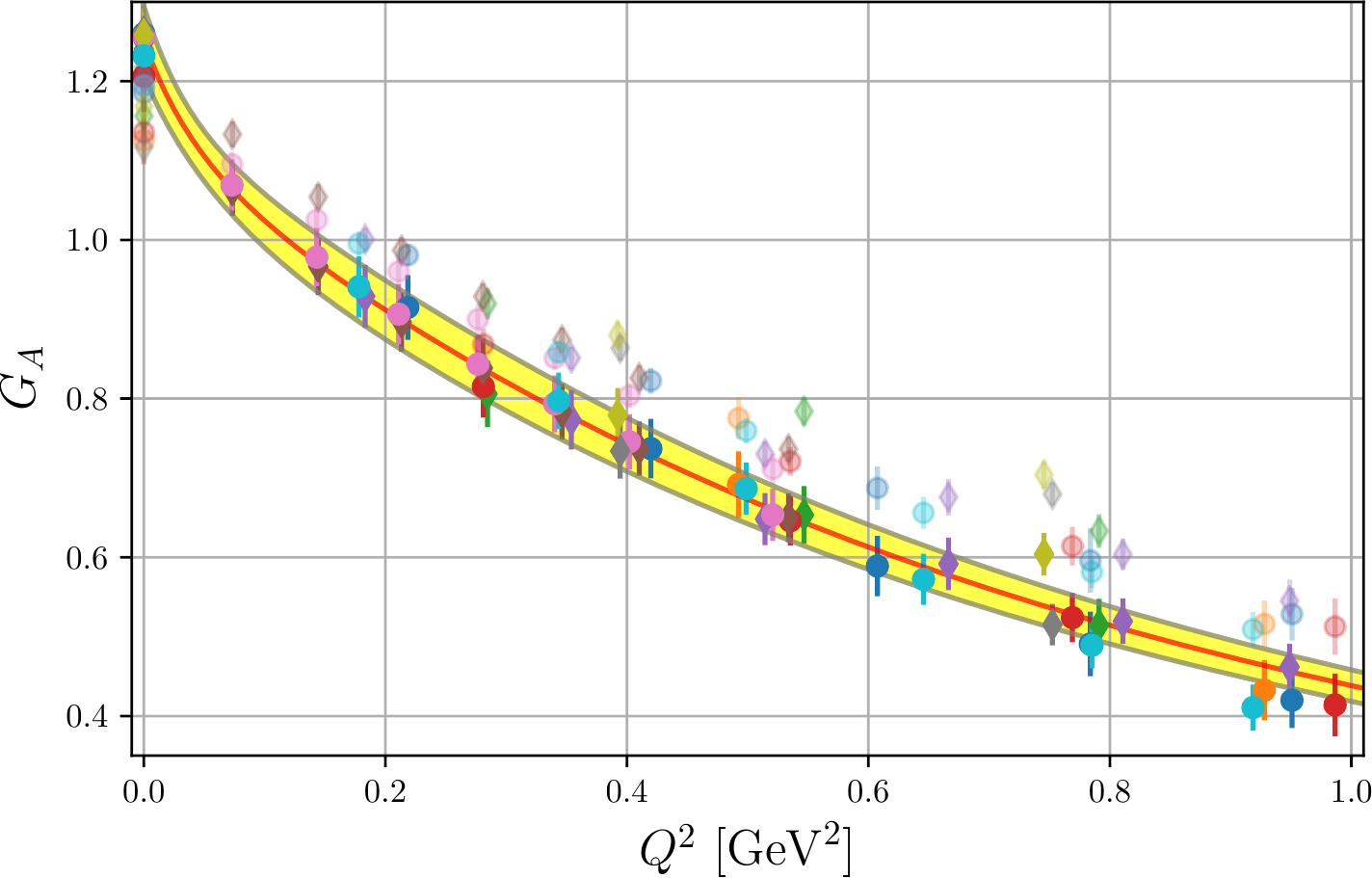}}\hfill
{\includegraphics[width=0.325\linewidth]{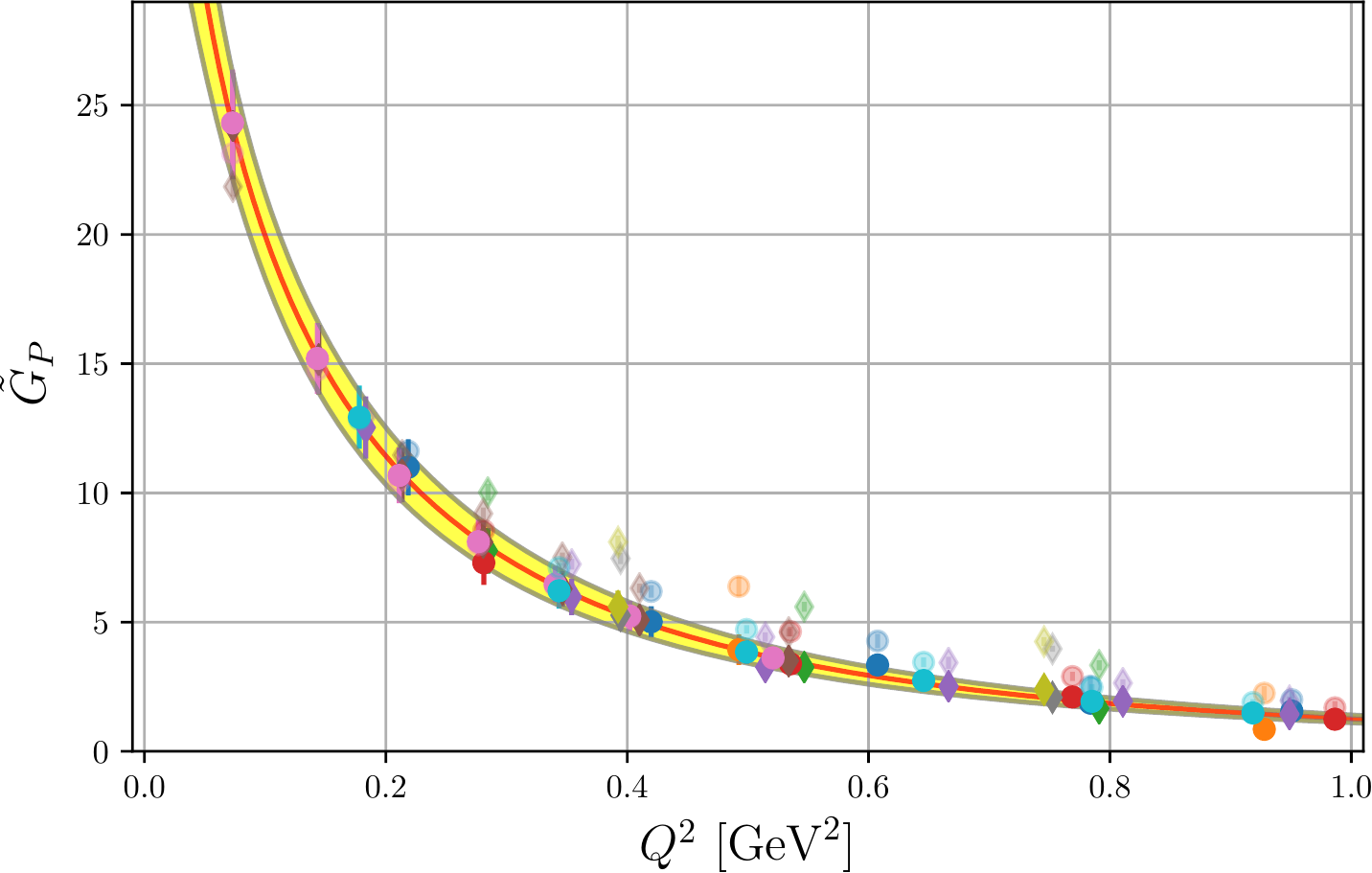}}\hfill
{\includegraphics[width=0.325\linewidth]{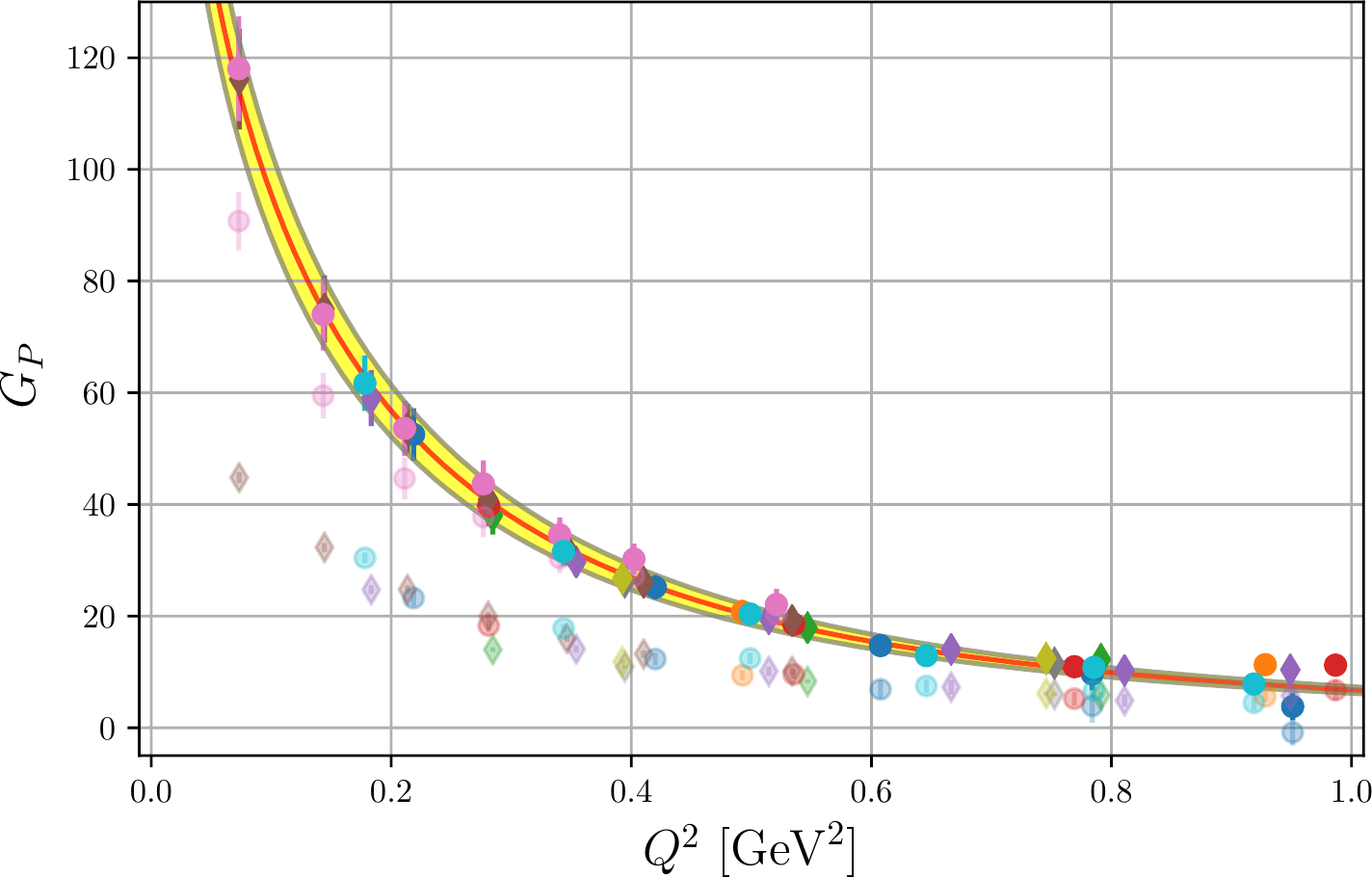}}
		\caption{\label{PlotFFs}Results for the form factors. Note that $G_P$ is scale-dependent; here it is plotted for the $\MSbar$ scale $\unit{2}{\giga\electronvolt}$. In the upper panels the results using the excited state subtraction explained in Section~\ref{sect_uncoverGS} (filled symbols) are compared to those obtained without this method (open symbols). The largest effect is found in the pseudoscalar form factor, while the others are almost not affected. In the lower panels the yellow band shows the final result for the form factors extrapolated to the physical point using the $z$-expansion described in Section~\ref{sect:parametrization}. In these plots the transparent points show the subtracted data itself, while the solid ones are parallel transported to the physical point in order to give an intuitive grasp on how good these fits actually describe the data points. The color coding follows Fig.~\ref{Fig:VolumeVSPion}.}
	\end{figure*}

One possible (and up to date the most widely-used) workaround to this problem is the exclusion of $A_0$ from the analysis.\footnote{For comparison we will show results obtained using this method in Figs.~\mbox{\ref{PlotRatiosPCAC}--\ref{PlotFFs}}.} A more satisfactory approach is to consider the current%
	\begin{equation}
		A^\perp_\mu = \left( g_{\mu\nu} - \frac{\overline{p}_\mu \overline{p}_\nu}{\overline{p}^2} \right) A^\nu\,,
	\end{equation}%
	which fulfills $\pbar^\mu\!A_\mu^\perp = 0$ by construction. Due to Eq.~\eqref{PmuAmu} one can be sure that the subtraction only removes excited state contributions, while leaving the ground state contribution unchanged. For the new current the three-point function has the expected behaviour, as shown in Fig.~\ref{CompareG4G5}.\par
	The connection between the axial and pseudoscalar channels via the PCAC relation suggests that similar excited state contributions are present in the pseudoscalar channel and it is advantageous to construct the combination%
	\begin{equation}
		P^\perp = P - \frac{1}{2i\,m_q} \frac{\pbar^\mu \pbar^\nu}{\pbar^2} \partial_\mu A_\nu \,, \label{Eqs:PBarDefinition}
	\end{equation}%
	such that $\partial^\mu\!A_\mu^\perp = 2i\,m_q P^\perp$. Again, Eq.~\eqref{PmuAmu} ensures that the ground state matrix element is not affected by this subtraction. We use the PCAC mass $m_q$ obtained from baryon three-point functions, cf.\ the discussion in Section~\ref{sect_quark_mass}, which we found leads to smaller discretization effects compared to employing the PCAC mass from pion two-point functions. The effect of Eq.~\eqref{Eqs:PBarDefinition} is illustrated in the right panel of Fig.~\ref{CompareG4G5}: While the original data looked less conspicuous than for the $A_0$ channel, the resulting shift is very significant.\par%
	To conclude this section, we remark that the subtraction constructed above should not be viewed as an operator improvement, as it depends on the external momentum. Instead, one should interpret it as a method to systematically construct combinations of correlation functions that suffer less from excited states. In principle, this method can be used wherever an exact relation for the ground state matrix elements exists (in this case Eq.~\eqref{PmuAmu}). However, the analogous constraint for the vector current, $q^\mu \la N^{\smash{\vec p^\prime}}_{\smash{\sigma^\prime}} | V_\mu | N^{\vec p}_{\sigma}  \ra = 0$, is fulfilled almost exactly by the data such that the method does not lead to an improvement.\par%
\section{Results}\label{sect_RESULTS}
	\subsection{Restoration of PCAC on the form factor level}
	We define the ratio (cf.\ also Ref.~\cite{Rajan:2017lxk})%
	\begin{align}
	 r_{\rm{PCAC}} &= \frac{m_q G_P (Q^2)+ \frac{Q^2}{4m_N} \tilde G_P(Q^2)}{m_N G_A(Q^2)} \,, \label{eq:ratioPCAC}
	\end{align}%
	where deviations from $r_{\rm{PCAC}}=1$ quantify the violation of the \PCACFF\ relation~\eqref{Eqs:NucleonFFPCAC}. Fig.~\ref{PlotRatiosPCAC} demonstrates that using the method described in Sect.~\ref{sect_uncoverGS} all ensembles, in particular the ones with small pion mass that previously exhibited the largest deviations, now fulfill the \PCACFF\ relation reasonably well. Even for large $Q^2\approx\unit{1}{\giga\electronvolt\squared}$ we see a significant improvement, although small deviations of~$\mathord{\sim}5\%$ remain. This residual violation can be attributed to $\mathcal O(a^2)$ discretization effects of Eq.~\eqref{Eqs:PCACRelation}.\par%
	In absence of better information, the induced pseudoscalar form factor is often estimated by%
	\begin{align}
	\tilde G_P &\overset{?}{\approx} \frac{4 m_N^2 G_A}{m_\pi^2+Q^2} & &\Rightarrow & r_{\rm{PPD} } &= \frac{(m_\pi^2+Q^2) \tilde G_P(Q^2)}{4 m_N^2 G_A(Q^2)} \overset{?}{=} 1 \,,\label{Eq:PPD}
	\end{align}
	usually called the pion pole dominance (PPD) assumption. Fig.~\ref{PlotRatiosPPD} demonstrates that this approximation does not describe the data, especially not at small $Q^2$. This is true for both, the original and the improved data (which reaffirms the findings of Refs.~\cite{Bali:2014nma,Rajan:2017lxk,Ishikawa:2018rew}). Hence, the observed disagreement with the PPD ansatz is certainly not caused by the same excited state effects that have been responsible for the \PCACFF\ violation. Since all data for~$r_{\rm PPD}$ collapse onto an almost universal function of $m_\pi^2+Q^2$ it seems highly unlikely that the deviation is due to discretization, volume, or quark mass effects.\par%
	\subsection{Parametrization of the form factors}\label{sect:parametrization}
	We parametrize the form factors using the $z$-expan\-sion~\cite{Hill:2010yb,Bhattacharya:2011ah}, which automatically imposes analyticity constraints. This corresponds to an expansion of the form factors in the variable%
	\begin{align}
	 z &= \frac{\sqrt{t_{\rm{cut}}+Q^2}-\sqrt{t_{\rm{cut}}-t_0\vphantom{Q^2}}}{\sqrt{t_{\rm{cut}}+Q^2}+\sqrt{t_{\rm{cut}}-t_0\vphantom{Q^2}}} \,, \label{Eqs:zDefinition}
	\end{align}
	where $t_{\rm{cut}}=9 m_\pi^2$ is the particle production threshold and $t_0$ is a tunable parameter.\footnote{Varying $t_0$ between $0$ and $t_{\rm{cut}}/2$ has no significant impact on the result. Therefore we have simply set it to zero in our analysis.} Isolating the pion pole in the pseudoscalar channels (cf.\ Ref.~\cite{Green:2017keo}) one obtains%
	\begin{align}
	 G_A &= \sum_{n=0}^N a^A_n z^n \,, &
	 \tilde G_P&= \frac{1}{m_\pi^2+Q^2}\sum_{n=0}^N a^{\tilde P}_n z^n \,, \label{Eq:zExpansionA}\\
	 &&G_P &= \frac{1}{m_\pi^2+Q^2}\sum_{n=0}^N a^{P}_n z^n \,.\label{Eq:zExpansionP}
	\end{align}
	To enforce the correct scaling in the asymptotic limit, $G_A \propto 1/Q^4$, $\tilde G_P \propto 1/Q^6$, and $G_P \propto 1/Q^6$~\cite{Alabiso:1974ye}, one has to implement the four constraints%
	\begin{align}
	 0&=\sum_{n=0}^N n^k a^X_n \,, \text{ for } k=0,1,2,3 \,,
	\end{align}
	which can be incorporated by fixing the first four coefficients according to
	\begin{align}
	 a^X_k &= \frac{(-1)^k}{k!(3-k)!} \sum_{n=4}^N \frac{n! }{(n-4)!(n-k)} a^X_n \label{Eqs:ZExpConstraints}
	\end{align}
	for $k=0,1,2,3$, such that one is left with $N-3$ free coefficients. In the following we will show the fit results with $3$ free coefficients (called a $z^{3+4}$ fit in the literature), since the $z^{2+4}$ fits failed to describe the data at low momentum transfer, while $z^{4+4}$ fits did not yield further improvement. In order to extrapolate to the physical point ($m_\pi\to m_\pi^{\text{phys}}$, $a\to0$, $L\to\infty$), we use the parametrization
	\begin{align}
	a^X_n &= b^X_n + c^X_n a^2 + d^X_n m_\pi^2 + e^X_n m_\pi^4 + f^X_n m_\pi^2 \frac{e^{-m_\pi L}}{\sqrt{m_\pi L}} \,.
	\end{align}
	This allows us to perform a combined fit to all ensembles with $15$ fit parameters for each form factor.\par%
	\subsection{Form factors, charges, and radii}\label{sect:chargesradii}
	Our results for the form factors are shown in Fig.~\ref{PlotFFs}, where the band shows the value extrapolated to the physical point using the combined fit to all ensembles described in the previous section. Since we know that the \PCACFF\ relation is fulfilled, we use $G_P(0)=G_A(0)m_N/m_q$ to obtain a data point for~$G_P$ in the forward limit (for each ensemble) in order to stabilize the continuum extrapolation of this form factor. In Table~\ref{Tab:ExpansionParameters} we list the $z$-expansion coefficients corresponding to our central values at the physical point.\par
	From the form factors we obtain the charges and the corresponding mean squared radii $r^2=-6 G^\prime(0)/G(0)$ as%
\begin{align}
          g_A = G_A(0) &= 1.25(4) \,, & r_A^2          &= \unit{0.79(15)}{\femto\meter\squared} \,,\notag\\
 \smash{\tilde G_P}(0) &= 64(23)  \,, & r_{\tilde P}^2 &= \unit{7.9(4.4)}{\femto\meter\squared} \,,\notag\\
          g_P = G_P(0) &= 280(27) \,, & r_P^2          &= \unit{7.3(1.0)}{\femto\meter\squared} \,.
\end{align}
The value of~$g_A$ is in perfect agreement with the experimental result $g_A/g_V=1.2724(23)$~\cite{Tanabashi:2018oca}. Our relatively large result for the squared axial radius still agrees within errors with $z$\nobreakdash-expansion fits to experimental $\nu d$~\cite{Meyer:2016oeg} and muon capture~\cite{Hill:2017wgb} data, but lies higher than other lattice results in the range $\unit{0.2-0.4}{\femto\meter\squared}$~\mbox{\cite{Green:2017keo,Alexandrou:2017hac,Capitani:2017qpc,Rajan:2017lxk}}, cf.\ Fig.~7 in Ref.~\cite{Hill:2017wgb}. We remark that our data are also well described by dipole fits, which result in significantly smaller radii.\par%
For the induced pseudoscalar coupling defined at the muon capture point~\cite{Bernard:2001rs} we obtain%
\begin{align} \label{gPt_result}
 \tilde g_P &= \frac{m_\mu}{2 m_N} \tilde G_P(0.88 m_\mu^2) = 2.8(7) \,,
\end{align}
where $m_\mu=\unit{105.6}{\mega\electronvolt}$ is the muon mass. Note that the small value obtained for $\tilde g_P$ is consistent with the strong violation of the pion pole dominance assumption at small momentum transfer shown in Fig.~\ref{PlotRatiosPPD}. The experimental value $\tilde g_P=8.06(48)(28)$ from muon capture~\cite{Andreev:2012fj} is consistent with PPD, but not with our data.\par%
\begin{table}[t]%
	\centering
	\caption{\label{Tab:ExpansionParameters}Parameters for the $z$-expansions with $t_0=0$ (cf.\ Eqs.~\eqref{Eqs:zDefinition} to~\eqref{Eq:zExpansionP}) representing our mean results for the nucleon form factors at the physical point. For completeness, we also give the values for $n=0,1,2,3$, which are fixed via Eq.~\eqref{Eqs:ZExpConstraints}.}%
	\begin{widetable}{\columnwidth}{cD{.}{.}{3.2}D{.}{.}{3.2}D{.}{.}{4.2}}%
        \toprule
        \multicolumn{1}{c}{$n$} & \multicolumn{1}{c}{$\hphantom{-}a^A_n$} & \multicolumn{1}{c}{$\hphantom{-}a^{\tilde P}_n\,[\giga\electronvolt\squared]$} & \multicolumn{1}{c}{$\hphantom{-}a^P_n\,[\giga\electronvolt\squared]$}\\
        \midrule
        0 &       1.25 &       1.17 &       5.10 \\
        1 &      -2.78 &      16.20 &      78.54 \\
        2 &      11.59 &     -58.27 &    -231.62 \\
        3 &     -43.58 &      47.61 &      39.03 \\
        4 &      70.23 &      27.18 &     410.70 \\
        5 &     -49.69 &     -52.15 &    -435.32 \\
        6 &      12.98 &      18.26 &     133.57 \\
        \bottomrule
	\end{widetable}%
\end{table}%
\section{Summary}%
We have presented a method to identify (and subtract) excited state contributions that spoil the PCAC relation on the form factor level. This mainly affects correlation functions involving the $P$ and $A_0$ currents, which have much larger coupling to the pion at small momentun transfer than~$A_i$. After our subtraction, \PCACFF\ is fulfilled up to small deviations at large momentum transfer, which can be interpreted as lattice artifacts. In spite of this improvement, we find that the pion pole dominance assumption that relates the axial and the induced pseudoscalar form factor is still strongly broken at small momentum transfer, which confirms the findings of Refs.~\cite{Bali:2014nma,Rajan:2017lxk,Ishikawa:2018rew}. A recent calculation within chiral perturbation theory~\cite{Bar:2018akl} (along the lines of Refs.~\cite{Bar:2017kxh,Bar:2017gqh} using interpolating currents from Ref.~\cite{Wein:2011ix}, cf.\ also Refs.~\cite{Tiburzi:2015tta,Tiburzi:2015sra}) indicates that this deviation at small momentum transfer might be due to additional large excited state contributions in the induced pseudoscalar form factor, which is almost unaffected by the subtraction method described in Section~\ref{sect_uncoverGS}. However, our data do not indicate any presence of these excited states.\par%
We parametrize the form factors using the $z$-expansion and extrapolate them to the physical point by a combined fit to all ensembles at hand. We find that these fits provide a very reasonable description of the data and we use them to extract the values of the charges $g_A$ and $g_P$. The result for the axial form factor exhibits a rather steep slope, which is not related to our improvement, but a consequence of the applied $z^{3+4}$~parametrization. It corresponds to an axial dipole mass $M_A = \sqrt{12}/r_A = \unit{0.77(8)}{\giga\electronvolt}$ that is in rough agreement with phenomenological values around $\unit{1}{\giga\electronvolt}$. The disagreement of our result for the induced pseudoscalar charge~$\tilde g_P$ with experiment could point to persistent problems in the form factor $\tilde G_P$ and warrants further investigation. All our physical results (extrapolated to physical quark masses, to infinite volume, and to the continuum) currently come with relatively large errors that are mainly caused by the narrow range of available lattice spacings. Reducing the systematic error at the physical point (e.g., by the analysis of CLS ensembles with finer lattice spacings~\cite{Bruno:2014jqa,Bali:2016umi}) will be of high priority in the future.\par%
\section*{Acknowledgements}
MG, PW, and TW are grateful to F.~Hutzler, while GB and SC enjoyed fruitful discussions with R.~Gupta and K.-F.~Liu. Many of the ensembles used in this work have been generated on the QPACE supercomputer~\cite{Baier:2009yq,Nakamura:2011cd}, which was built as part of the Deutsche Forschungsgemeinschaft (DFG) Collaborative Research Centre/Trans\-regio~55 (SFB/TRR-55). The analyses were performed on the iDataCool cluster in Regensburg and the SuperMUC system of the Leibniz Supercomputing Centre in Munich.\par%
\FloatBarrier

\end{document}